# SUPERCONDUCTIVITY IN $SmFe_{1-x}Co_xAsO$ (x = 0.0 to 0.30)

V.P.S. Awana*, Anand Pal, Arpita Vajpayee, R.S. Meena, and H. Kishan
National Physical Laboratory, Dr. K.S. Krishnan Marg, New Delhi 110012, India

Mushahid Husain
Department of Physics, Jamia Millia Islamia, New Delhi-110025, India

R. Zeng

ISEM, University of Wollongong, NSW 2522, Australia

S.Yu[1,2], K. Yamaura[1,2], and E. Takayama-Muromachi[1,2,3]
[1]JST, Transformative Research-Project on Iron Pnictides (TRIP), Tsukuba, Ibaraki 305-0044, Japan
[2]Superconducting Materials Center, National Institute for Materials Science, 1-1 Namiki, Tsukuba, Ibaraki 305-0044, Japan
[3]International Center for Materials Nanoarchitectonics (MANA), National Institute for Materials Science, Tsukuba, Ibaraki 305-0044, Japan

* Corresponding Author

Dr. V.P.S. Awana

National Physical Laboratory, Dr. K.S. Krishnan Marg, New Delhi-110012, India

Fax No. 0091-11-45609310: Phone No. 0091-11-45608329

e-mail-awana@mail.nplindia.ernet.in: www.freewebs.com/vpsawana/

**Abstract**

We report synthesis, structural details and magnetization of $SmFe_{1-x}Co_xAsO$ with x ranging from 0.0 to 0.30. It is found that Co substitutes fully at Fe site in SmFeAsO in an iso-structural lattice with slightly compressed cell. The parent compound exhibited known spin density wave (SDW) character below at around 140 K. Successive doping of Co at Fe site suppressed the SDW transition for x = 0.05 and later induced superconductivity for x = 0.10, 0.15 and 0.20 respectively at 14, 15.5 and 9K. The lower critical field as seen from magnetization measurements is below 200Oe. The appearance of bulk superconductivity is established by wide open isothermal magnetization M(H) loops. Superconductivity is not observed for higher content of Co i.e. $x \geq 0.30$. Clearly the Co substitution at Fe site in $SmFe_{1-x}Co_xAsO$ diminishes the Fe SDW character, introduces bulk superconductivity for x between 0.10 and 0.20 and finally becomes non-superconducting for x above 0.20. The $Fe^{2+}$ site $Co^{3+}$ substitution injects mobile electrons to the system and superconductivity appears, however direct substitution introduces simultaneous disorder in superconducting FeAs layer and thus superconductivity disappears for higher content of Co.

## 1. Introduction

Soon after the discovery of superconductivity at 26 K in $LaO_{1-x}F_xFeAs$, the substitution of La by other rare earth elements such as Ce, Pr, Sm, Nd, Gd and Tb has led to a family of 1111 phase high-$T_c$ superconductors [1-9]. The parent compound REFeAsO (RE= rare earth) itself is non-superconducting but shows an anomaly (spin density wave) at around 150 K. Doping of fluorine into insulating REO layer provides the electrons as charge carriers in FeAs layer, which suppresses the anomaly and superconductivity is obtained. Carrier doping plays a major role in the appearance of superconductivity by suppressing the magnetic order in these complex compounds. Recently it has been shown that superconductivity can be induced in LaFeAsO, CeFeAsO and SmFeAsO by cobalt substitution [10-13] in place of iron that results in direct injection of electrons in the conducting FeAs layers.

Here we report the synthesis and characterization of Co-doped SmFeAsO and obtained superconductivity for 0-20% doping levels. Chemically, Co is a better mean of electron doping and carriers are doped directly into the FeAs planes [10-13]. For higher Co content the superconductivity disappears. This is most probably due to over doping.

## 2. Experimental Details

Polycrystalline $SmFe_{1-x}Co_xAsO$; x=0.0, 0.15, 0.20 & 0.30 samples were synthesized by single step solid-state reaction method. Stoichiometric amounts of Sm, Fe, As, $Co_3O_4$ and $Fe_2O_3$ were thoroughly ground. It is to be noted that weighing and grinding was done in the glove box under high purity argon atmosphere. The powder was then palletized and vacuum-sealed in a quartz tube. Roughly the vacuum was of the order of $10^{-4}$ Torr. Subsequently this sealed quartz ampoule was placed in box furnace and heat treated at 550$^{o}$C for 12 hours, 850$^{o}$C for 12 hours and then finally 1150$^{o}$C for 33 hours in continuum. Then furnace was allowed to cool naturally.

The X-ray diffraction pattern of the compound was taken on Rigaku mini-flex II diffractometer. The resistivity measurements were carried out by conventional four-probe method on a quantum design *PPMS* (Physical Property Measurement System). Heat capacity measurements Cp(T) were also carried out on the same *PPMS*. The magnetization measurements were carried out on Quantum Design magnetic property measurement system (*MPMS*).

## 3. Results and Discussion

The room temperature X-ray diffraction (XRD) pattern for $SmFe_{1-x}Co_xAsO$ samples with x=0.0, 0.15, 0.20 & 0.30 along with their Rietveld analysis are shown in Fig.l(a). The structure of $SmFe_{1-x}Co_xAsO$ at room temperature for all compositions of x was refined with the tetragonal space group *P*4/*nmm* (space-group no.129). The Sm and As both atoms are located at Wyckoff positions *2c*, O is situated at *2a* and Fe/Co are shared at site *2b*. Good fits to the reported structural model are observed. It is observed that all main peaks can be well indexed based on the space group *P*4/*nmm*, indicating that the studied samples are essentially single phase. Some extra peaks at around 30 and 43 degree having very low intensity are also seen in the XRD pattern of pure SmFeAsO and $SmFe_{0.85}Co_{0.15}AsO$ sample, which are due to the minor presence of SmAs. Another small peak at around 34.5 degree shows the minute presence of CoAs for $SmFe_{0.80}Co_{0.20}AsO$ and $SmFe_{0.70}Co_{0.30}AsO$ sample. The exact final compositions of all the studied samples are not determined. Infact, compositional analysis techniques such as EPMA/EDAX are highly desired on these samples. Detailed compositional, structural and micro-structural analysis on these samples is already underway and will be reported later.

The Lattice parameters for $SmFe_{0.9}Co_{0.1}AsO$ are found to be *a* = 3.9398 (7) Å and *c* = 8.4639 (4) Å, while for the parent compound the lattice parameters are *a* = 3.9375 (6) Å and *c* = 8.5021 (4) Å. Clearly Co doping leads to decrement in the c-axis lattice while the a-axis remain nearly unaltered {Fig.1(b)}. The decrease in *c* lattice parameter suggests towards the successful chemical substitution of Co in the tri-valence $Co^{3+}$ state at the place of $Fe^{2+}$. The shrinkage of c-axis suggests the strengthening of interlayer Coulomb attraction, implying the increase of density of negative charge in FeAs layers by the Co doping. The incorporation of Co in the Fe site reduces the cell volume due to contraction of the c-axis lattice constant.

Figure 2(a) shows the temperature dependence of the DC magnetization measured under zero-field-cooled (*ZFC*) and field-cooled (*FC*) conditions at 10Oe for $SmFe_{1-x}Co_xAsO$; x=0.10, 0.15, 0.20 samples. The susceptibility becomes negative (diamagnetic) below temperature 14, 15.5 and 9 K for x = 0.10, 0.15 and 0.20 respectively. This shows that x=0.15 possess the highest transition temperature ($T_c$) among the three superconducting samples. The $SmFe_{0.70}Co_{0.30}AsO$ does not show superconductivity in magnetization measurement. This is most probably due to over doping of carriers by $Fe^{2+}$ substitution with $Co^{3+}$. Another reason could be the extent of disorder due to the substitution of $Fe^{2+}$ by $Co^{3+}$ in FeAs superconducting layers. The magnitude of the diamagnetic signal confirms the bulk superconductivity in our samples. The inset of Fig 2(a) shows M versus H at T=2 K for all the three superconducting samples. The M(H) loop opening at 2K is maximum for x = 0.15 sample, coinciding with its highest $T_c$ of 15.5K.

Temperature dependence of DC susceptibility of highest $T_c$ $SmFe_{0.85}Co_{0.15}AsO$ sample is further shown in Fig. 2(b) separately and its magnetization loops at various temperatures of 9, 7, 5 & 2 K are depicted in insets of the same figure. Although transition is broad, the compound undergoes into superconducting state at below 15.5 K. Upper inset shows the first quadrant of M(H) loop at temperature 9, 7, 5 & 2 K for $SmFe_{0.85}Co_{0.15}AsO$, where linear response of the Meissner state up to less than 180 Oe is followed by a decrease in magnitude of M with increasing H, as is typical of flux penetration in the vortex state of a type-II superconductor. This gives the lower critical field ($Hc_1$) of around 180 and 50 Oe at 2 and 9K respectively. As expected the $Hc_1$ decreases with increase in temperature from 2 to 9K. Lower inset shows the complete M(H) loops for $SmFe_{0.85}Co_{0.15}AsO$ at 2, 5 and 9K in applied fields of 2000 Oe to -2000 Oe. Although wide open loops are seen right up to 9K, but with a positive background, which arise from para-magnetic contributions from $Sm^{3+}$, $Fe^{2+}$ and $Co^{3+}$ spins.

The temperature dependence of electrical resistivity ρ(T) in zero field is shown in main panel of Fig. 3 for $SmFe_{0.85}Co_{0.15}AsO$ sample and inset of Fig. 3 shows the variation of resistivity under magnetic field ρ(T)H up to 9 Tesla field. At room temperature the resistivity value is 3.8 mΩ cm and it decreases with temperature as expected for metallic behavior. Finally at 14 K the $SmFe_{0.85}Co_{0.15}AsO$ sample becomes superconducting. The superconducting transition temperature being determined from resistivity measurements as ρ=0 is at 14 K; though the diamagnetic transition temperature ($T_c^{dia}$) is slightly higher at 15.5 K, see Fig. 2(a). This is because $T_c^{dia}$ coincides roughly with the superconducting onset. The parent compound SmFeAsO does not show superconductivity and the same rather exhibits the SDW characteristic at around 140 K [14]. ρ(T)H measurements reveal that the resistive transitions for this superconducting composition shift to lower temperatures by applying a magnetic field. The transition width for the sample becomes wider with increasing H. If we focus on the temperature at which resistivity becomes negligible ($\rho \rightarrow 0$) we find that rate of decrease of transition temperature with applied magnetic field is 1 Kelvin per Tesla {$dT_c/dH \sim 1K/T$} for Co substituted $SmFe_{0.85}Co_{0.15}AsO$ sample. This value is far less than that of YBCO {$dT_c/dH \sim 4K/T$} and $MgB_2$ {$dT_c/dH \sim 2K/T$} samples; it suggests toward a high value of upper critical field ($H_{c2}$) in these compounds [15]. We can say that oxypnictides are emerging as a new class of high field superconductors surpassing the $H_{c2}$ of $Nb_3Sn$, and $MgB_2$. In fact FeAs seems surpassing the 100 Tesla field benchmarks of the high-$T_c$ cuprate superconductors.

We observed that by doping the magnetic ion $Co^{3+}$ in the conducting Fe-As plane of SmFeAsO, the SDW state of $Fe^{2+}$ is suppressed and superconductivity is induced by induction of mobile electron carriers in the system. This is unlike the cuprate superconductors, in which substitution of Cu with its neighbors in the periodic table (Ni, Zn, Co, and Fe etc.) in $CuO_2$ plane severely destroys the superconductivity [16].

## 4. Conclusions

We have successfully synthesized the iron-based Co-doped layered compound with nominal composition $SmFe_{1-x}Co_xAsO$; x=0.0 to 0.30 by one-step solid-state reaction method at normal pressure. Cobalt acts as an effective dopant and produces superconductivity in this system at 14 K for $SmFe_{0.90}Co_{0.10}AsO$ sample. The resistive transition in fields suggests that $H_{c2}$ is remarkably high, which indicates encouraging potential applications.

## Acknowledgements

The work is supported by Indo-Japan (DST-JSPS) bilateral exchange research program. VPSA further thanks NIMS for providing him with the MANA visiting scientist position for three months. Authors from NPL would like to thank Prof. Vikram Kumar (DNPL) for his constant encouragement. Anand Pal and Arpita Vajpayee are thankful to CSIR for providing the financial support during their research.

**Figure Captions**

Figure 1(a): Fitted and observed room temperature X-ray diffraction patterns of $SmFe_{1-x}Co_xAsO$; x=0.0, 0.15, 0.20 & 0.30 compounds; impurity peaks are marked with '*' and '#' in the figure.

Figure 1(b): Effect of Co content on lattice parameters and unit cell volume

Figure 2(a): Temperature variation of magnetic susceptibility M(T) in *FC* and *ZFC* condition for $SmFe_{1-x}Co_xAsO$; x=0.10, 0.15 & 0.20 compounds. Inset shows complete magnetization loops M(H) at 2 K for the same.

Figure 2(b): Temperature dependence of magnetic susceptibility M(T) in *FC* and *ZFC* condition for $SmFe_{0.85}Co_{0.15}AsO$ sample. Inset (i) shows the complete magnetization loops M(H) at 2, 5 & 9 K for the same and inset (ii) shows the first quadrant of M(H) loop at 2, 5, 7 & 9 K for $SmFe_{0.85}Co_{0.15}AsO$ sample.

Figure 3: Resistivity behavior with temperature variation $\rho(T)$ of $SmFe_{0.85}Co_{0.15}AsO$ sample at zero field. Inset shows the variation of resistivity in the presence of applied magnetic field $\rho(T)H$ up to 9 Tesla.

Figure 1a

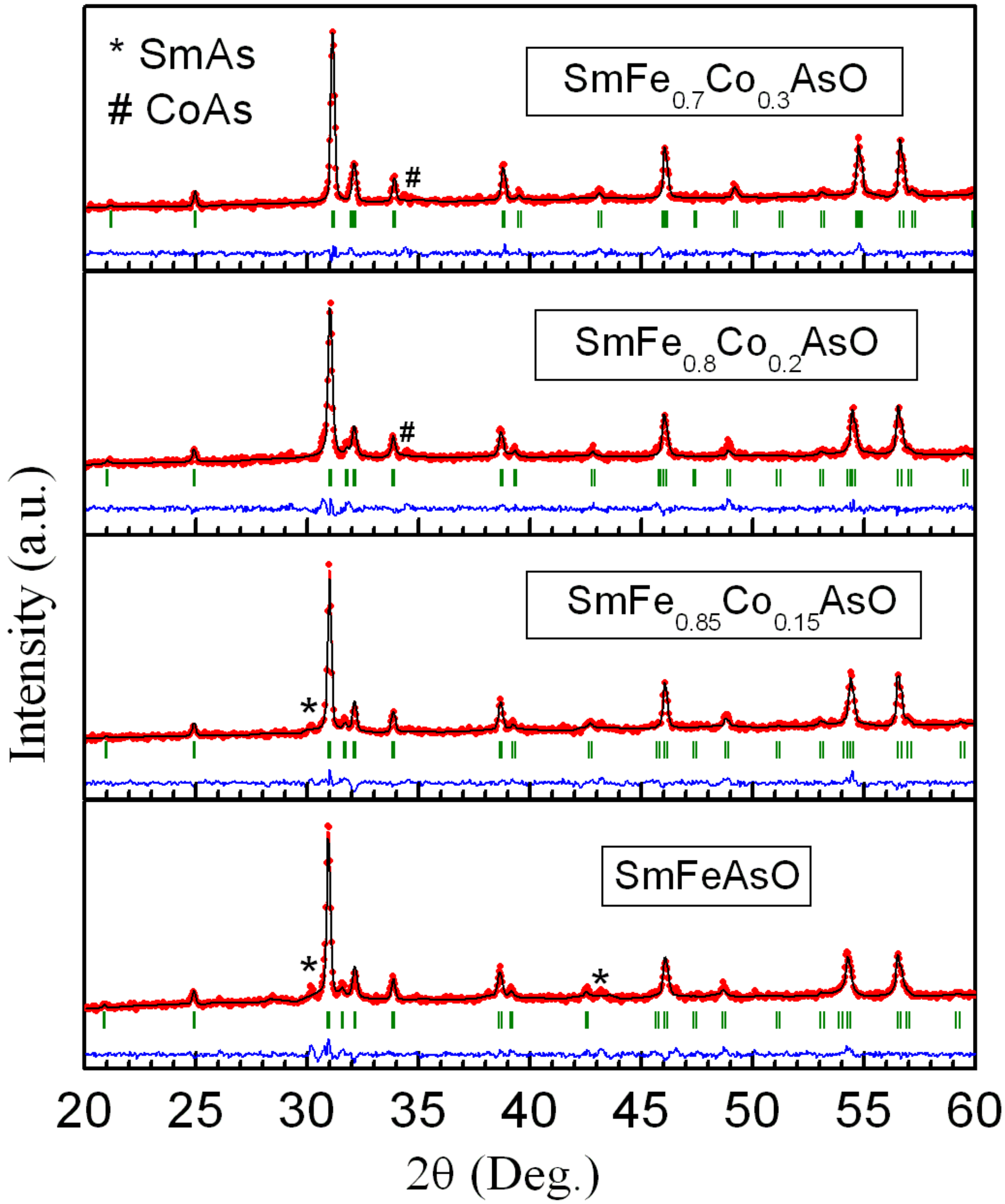

Figure 1b

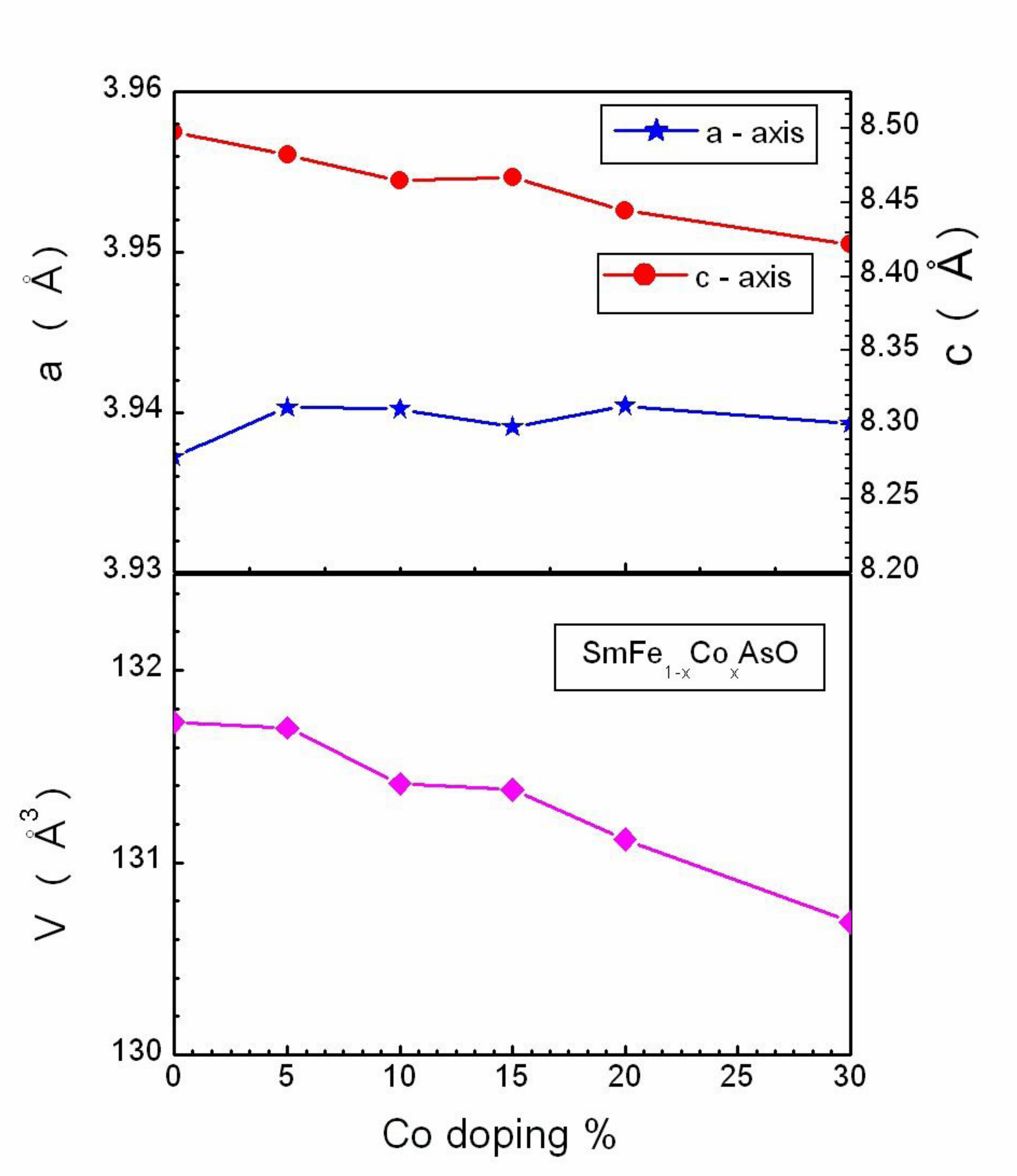

Figure 2a

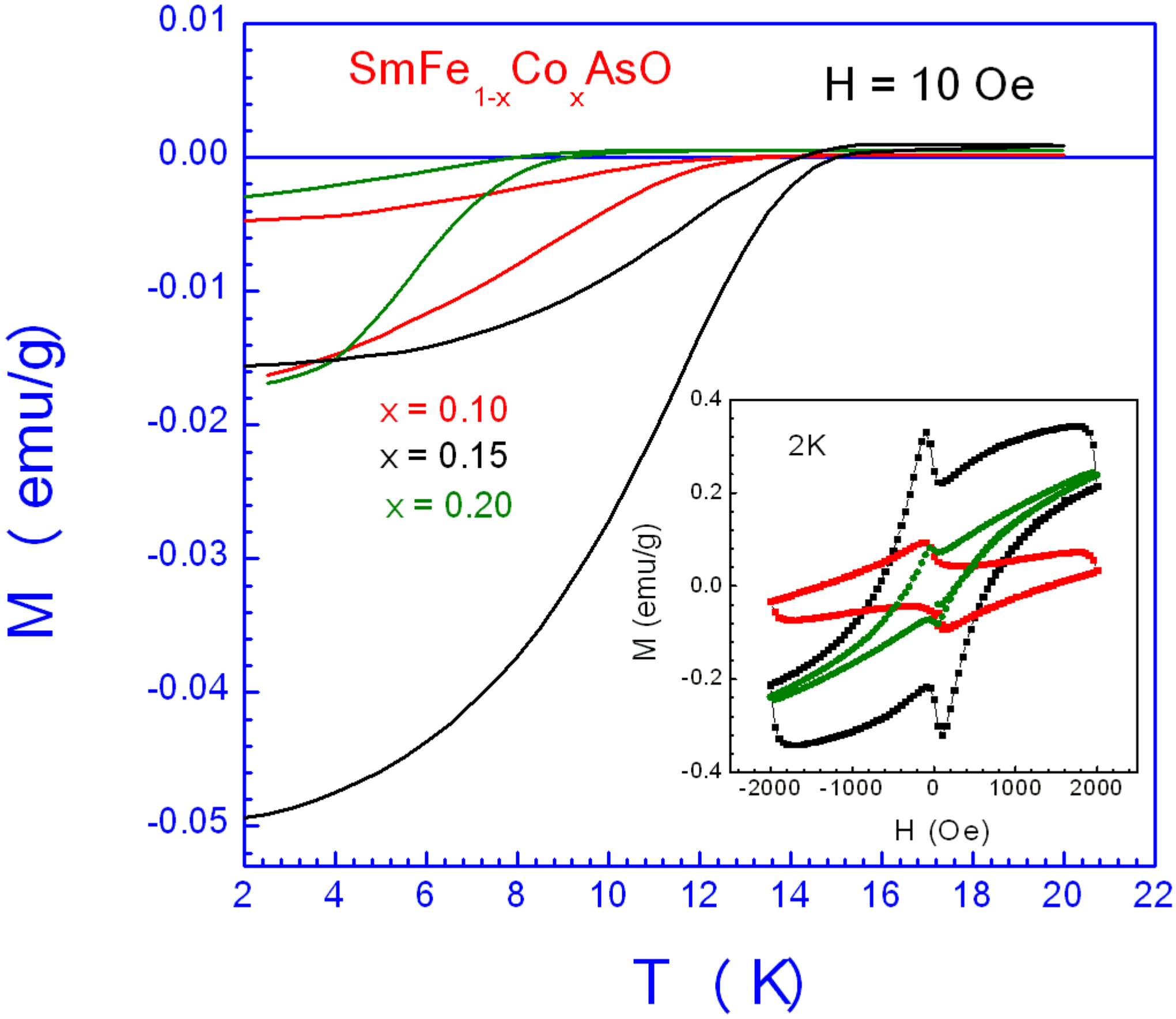

Figure 2b

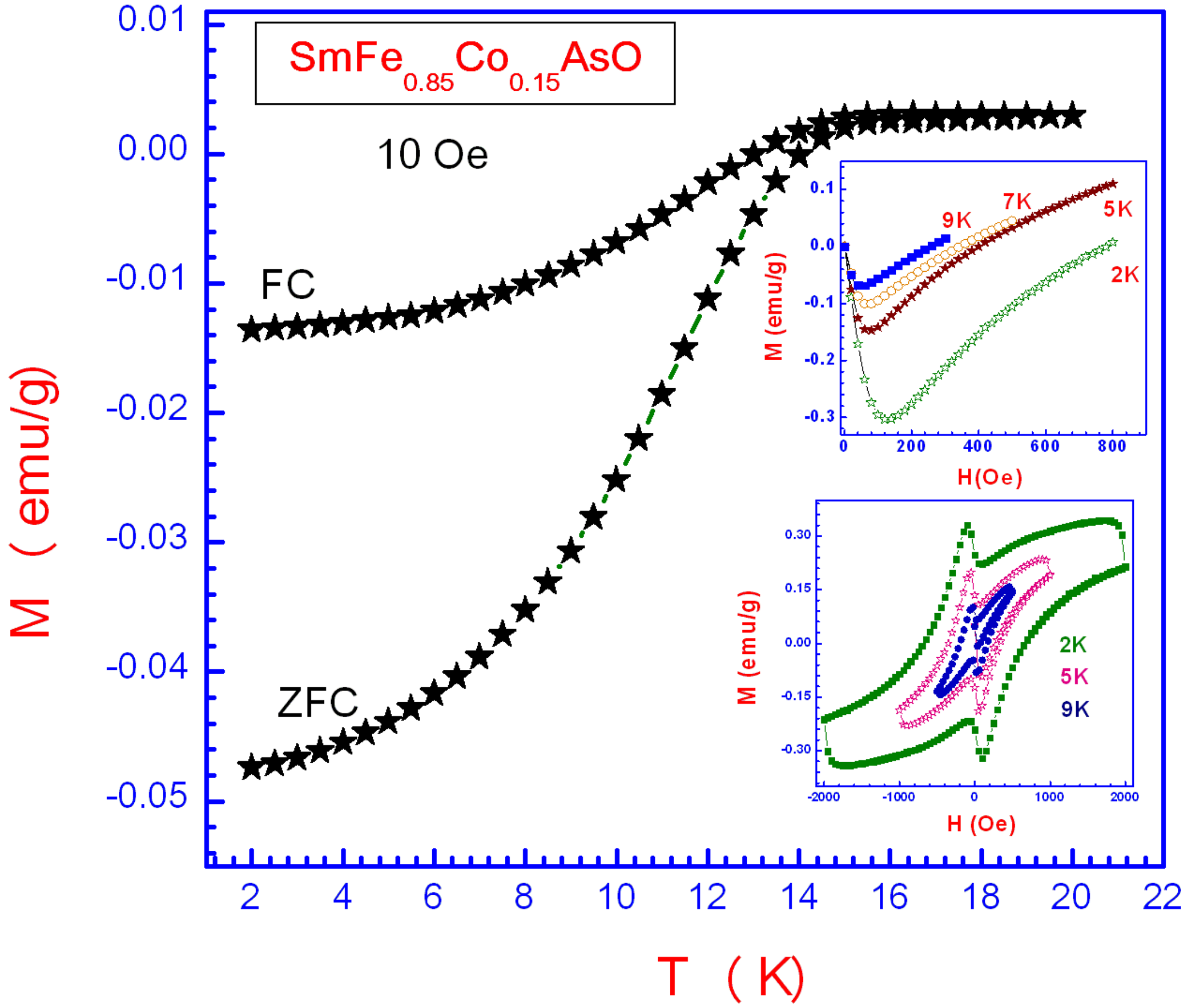

Figure 3

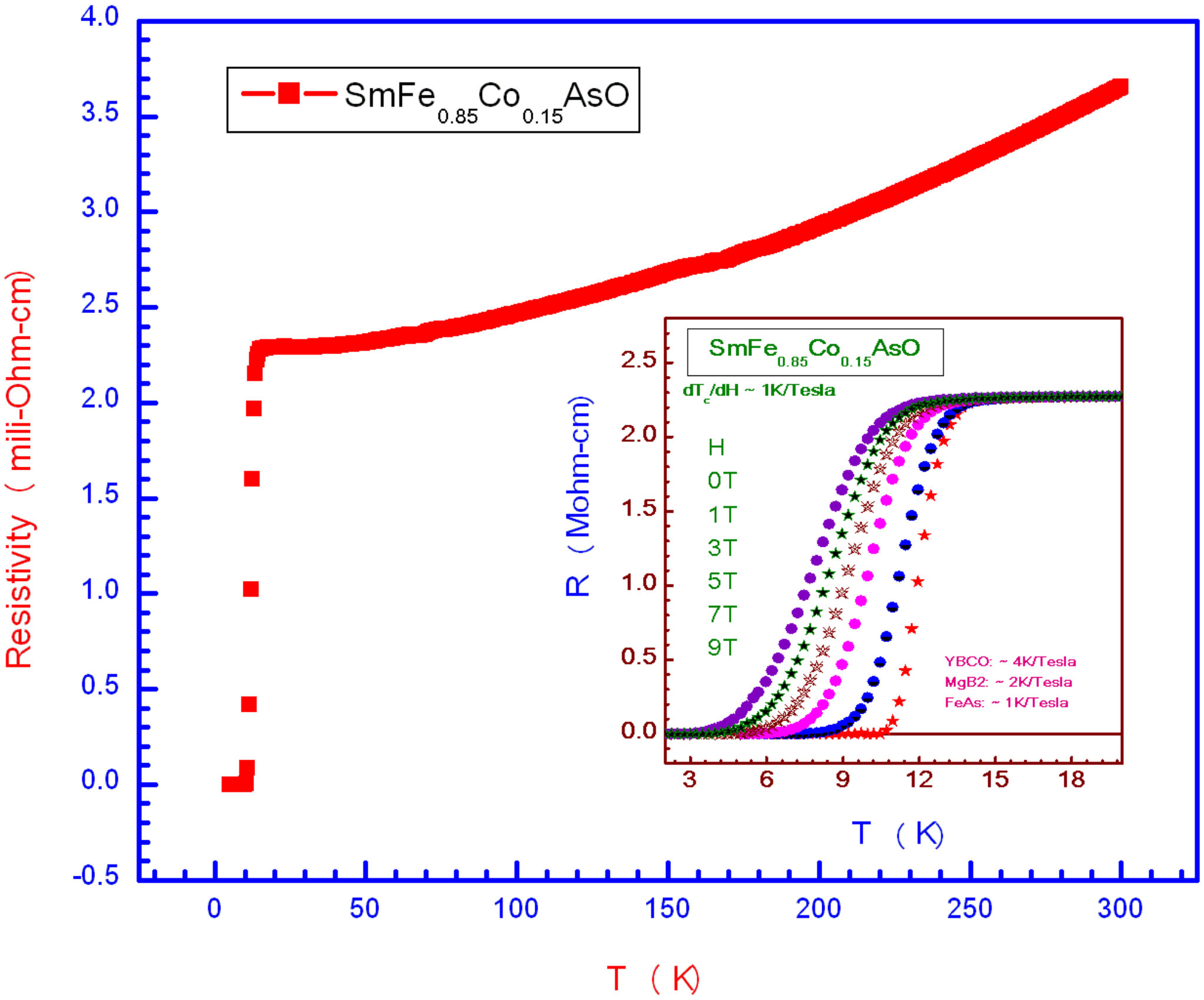